\documentclass[9pt,twocolumn]{extarticle}

\usepackage[T1]{fontenc}

\usepackage[margin=1.8cm]{geometry} 
\usepackage{graphicx}
\usepackage{amsmath, amssymb}
\usepackage{dcolumn}
\usepackage{bm}
\usepackage{float}
\usepackage{authblk}
\usepackage[numbers]{natbib}

\usepackage{siunitx}
\DeclareSIUnit{\sample}{S}

\title{\LARGE\bfseries Release and Recapture of Silica Nanoparticles from an Optical Trap in Weightlessness}

\author[1]{Govindarajan Prakash}
\author[1]{Sven Herrmann}
\author[2]{Ralf B. Bergmann}
\author[2]{Christian Vogt\thanks{vogt@bias.de}}

\affil[1]{Zentrum für angewandte Raumfahrttechnologie und Mikrogravitation (ZARM), Universität Bremen, Bremen, Germany}
\affil[2]{BIAS—Bremer Institut für Angewandte Strahltechnik GmbH, Klagenfurter Str. 5, 28359 Bremen, Germany}

\date{\today}
\begin{document}
\maketitle		
	\begin{abstract}
		\noindent
	 	Optically trapped Silica nanoparticles are a promising tool for precise sensing of gravitational or inertial forces and fundamental physics, including tests of quantum mechanics at 'large' mass scales. This field, called levitated optomechanics can greatly benefit from an application in weightlessness. In this paper we demonstrate the feasibility of such setups in a microgravity environment for the first time. Our experiment is operated in the GraviTower Bremen that provides up to \SI{2.5}{\second} of free fall. System performance and first release-recapture experiments, where the particle is no longer trapped are conducted in microgravity. This demonstration should also be seen in the wider context of preparing space missions on the topic of levitated optomechanics.
	\end{abstract}
	
	\section*{Introduction}
	
	The unification of the theory of relativity and quantum mechanics remains one of the biggest challenges in today's physics. For example the gravitational potential created by a massive particle in superposition is unknown \cite{Carlesso2019}. In order to solve this puzzle scientists approach this interplay from two sides. On the one hand, the gravitational field of ever smaller source masses is investigated \cite{Westphal2021}, on the other, quantum mechanical behavior in ever larger mass systems is observed. \newline
	An established approach to prove quantum mechanical behavior is to observe interference of a system. The current mass record of \SI{25}{\kilo\dalton} for such observation was done with molecules \cite{Fein2019}.\newline
	A promising candidate to observe this behavior for even larger masses is levitated optomechanics \cite{Millen2020, gonzalez-ballestero_levitodynamics_2021}, where Silica nanoparticles are trapped in an optical trap, formed by a focused laser beam. These systems are operated in vacuum chambers and can provide a high degree of isolation from environmental disturbances. \newline
	These properties are generally suitable for observing interference with relatively large masses, as interactions with the environment lead to decoherence and thus a smearing of the desired signal. In this context, the most important contributions to decoherence are collisions with gas molecules, as well as scattered, absorbed or emitted photons. In principle, it must be prevented that position information about the particle is available. While collisions can be prevented by an extremely good vacuum, an optical trap always provides a large number of photons that can be absorbed or scattered. It is therefore desirable to switch off the trap. Operation in zero gravity offers exactly this possibility, while the particle is still available for measurements for several seconds. These considerations led for example to the proposal of the MAQRO-mission, a dedicated satellite mission to test large mass particles for quantum mechanical behavior \cite{kaltenbaek_macroscopic_2012,kaltenbaek_research_2023}, including a launched pathfinder mission to demonstrate particle loading in space \cite{homans_experimental_2025} \newline
	Another aspect to the presented work here is the possibility for measuring tiny forces with free flying test particles. Hebestreit \textit{contain.} demonstrated the measurement of the Earth's gravitational acceleration with a free falling nanoparticle \cite{Hebestreit2018}. While in this experiment the particle was dragged out of the trapping region by gravity, our approach in weightlessness in principle allows for extended free fall times and increased force sensitivity.\newline
	Our experiment demonstrates for the first time the general feasibility of using levitated optomechanics in a weightlessness environment. In addition we demonstrate first release and recapture experiments paving the way to precise force measurements and 'large' mass interference.
    
	\section*{Experimental Setup}
\begin{figure}[b]
		\includegraphics[width=0.48\textwidth]{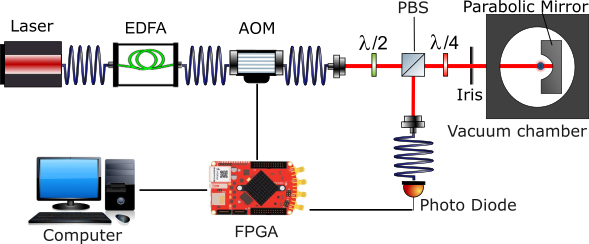}
		\caption{\textbf{Schematic drawing.} Experimental setup to optically trap Silica nanoparticles in vacuum. Detailed description in the text. }
		\label{fig:aufbauschemacropped}
	\end{figure}

    Figure \ref{fig:aufbauschemacropped} schematically shows our experimental design of an optical trap for transparent nanospheres in the focus of a laser while being operated in the weightlessness environment of the Drop Tower in Bremen. \newline
	Laser light at \SI{1550}{\nano\meter} with a power of approximately \SI{12}{\milli\watt} is generated by a fiber coupled laser diode (Laser). The linearly polarized light passes an erbium doped fiber amplifier (EDFA) providing approximately \SI{500}{\milli\watt} to be fed into an acousto-optic modulator (AOM) for intensity control. From here the light is guided free beam through a half wave plate ($\lambda/2$) and a polarization dependent beam splitter (PBS). Another quarter wave plate ($\lambda/4$) converts the polarization from linear to circular. The collimated beam passes through an iris, necessary for our alignment procedure \cite{Vovrosh2018}, and hits a high numerical aperture (NA) parabolic mirror inside a vacuum chamber. The incoming light is focused and forms an optical trap for Silica particles. The mirror is diamond turned from aluminum with an NA higher than 0.9. \newline
	Backscattered light from the particle is collimated by the mirror and passes the waveplate again leaving a linearly polarized beam with perpendicular polarization to the incoming beam. Therefore, the beam splitter allows to separate incoming and outgoing beams. The scattered light is coupled into a single mode fiber and guided to a single photodiode. \newline
	The photo diode provides an interferometric measurement between the scattered light, and a small fraction of the highly divergent field which passed the particle without scattering events \cite{Vovrosh2017a}. 
	Movement of the particle inside the optical trap translates to voltage differences on the photodetector, which are recorded on a digital oscilloscope running on a field programmable gate array (FPGA). This approach was preferred over more sophisticated detection schemes to align with the need for further compactification in future space missions. \newline
	Particles with a diameter of approximately \SI{140}{\nano \meter} are loaded into the trap by spraying them solved in ethanol with a nebulizer into the trapping region \cite{summers_trapping_2008}.
    
	\subsection*{The GraviTower Bremen Pro}
	The GraviTower Bremen Prototype is a second generation drop tower with a height of \SI{16} {meters}. With an initial acceleration of up to \SI{5}{\gram} it provides up to \SI{2.5}{\second} of weightlessness on a parabolic vertical trajectory \cite{raudonis_microgravity_2023}. In contrast to first generation drop towers, the new generation is actively driven to compensate for air drag. This eliminates the need for a vacuum environment and greatly reduces the experimental cycle by a factor of approximately 50. A sledge, running on Polytetrafluoroethylene (PTFE) skids is driven by ropes connected to a hydraulic wrench. Residual vibrations are further reduced by a free flying system inside the sledge decoupled by several air bearings. The initial acceleration can be chosen between \SI{1.5}{\gram}  and \SI{5}{\gram}. For the work in this paper the acceleration was fixed to \SI{1.5}{\gram} providing approximately \SI{1.5}{\second} of free flight. While this value is long enough for the performed experiments, the system was also capable of performing equally well with accelerations of up to 3.5 g, higher values were not tested. \newline
	The GraviTower sets certain requirements on our experiment. The maximum volume of \SI{0.85}{\cubic\meter} and a maximum weight of \SI{500}{\kilogram} can easily be met. The system must be mounted on Aluminum/Plywood platforms with approximately \SI{70}{\centi\meter} in diameter. Different platforms are connected via four aluminum stringers with a length of \SI{1.34}{\meter}  \cite{ZentrumfurangewandteRaumfahrtundMikrogravitation2009}. Furthermore, the setup must have autonomous power supply and run sequences with a single starting trigger. \newline
	
	\section*{Results}	
	The aim of our experiments is to demonstrate the feasibility for levitated optomechanical experiments in a microgravity environment. Therefore, trap frequency measurements at different laser powers were compared for flight and laboratory data and the particle's movement in free flight is investigated. \newline
	
	\subsubsection*{Trap frequency measurements}
	
\begin{figure}[t]
      \centering
		\includegraphics[width=0.48\textwidth]{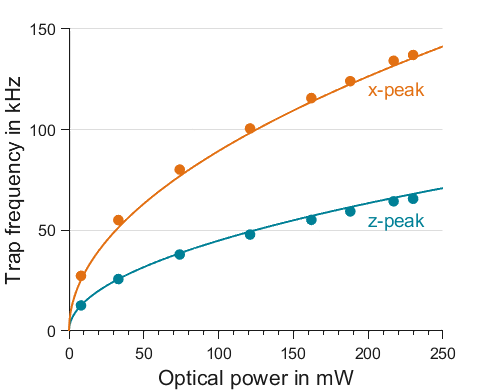}
		\caption{\textbf{Trap frequencies.} Measured in dependency of trapping power. Dots represent measurements while solid lines are fitted to Eq \ref{eq:trap_freq} leaving $\omega_0$ as free parameter.}
		\label{fig:trapfreq}
    \end{figure}

	Trap frequencies are resolved by a Fourier transformation of the oscillatory interference signal recorded at the photo diode. For these measurements the incoming light was circularly polarized leaving all radial frequencies to be degenerated.	No differences in trap frequencies are observed between the presence and absence of gravity. This is in accordance with numerical simulations of the particle's movement in the trap, based on a second order symplectic integrator \cite{Donnelly2005}. \newline
    Figure \ref{fig:trapfreq} shows the data in dependence of optical trapping power.  
    The equations for radial ($\omega_r$) and axial trap frequency ($\omega_{ax}$) are
	\begin{equation}
		\omega_r=\sqrt{\frac{4\alpha P_0}{m\pi c \epsilon_0 \omega_0^4}} \quad\text{and}\quad  \omega_{ax}=\sqrt{\frac{2\alpha \lambda^2 P_0}{m\pi^3 c \epsilon_0 \omega_0^6}}
		\label{eq:trap_freq}
	\end{equation}
	where $\alpha$ and m are the real part of the polarizability of the particle and it's mass, $c$ is the speed of light, $\epsilon_0$ the electric constant and $P_0$ the optical trapping power.	  
	Fitting equation \ref{eq:trap_freq} to the data reveals a beam waist of \mbox{$\omega_0$ = \SI{0.7}{\micro\meter}} meaning an effective numerical aperture of approximately 0.7. The discrepancy to the expected value seems to be a result of an improper alignment procedure \cite{Vovrosh2018} and could not be fixed during the flight campaign. 
    The iris just in front of the parabolic mirror, was closed further than needed and therefore blocked some outer parts of the parabolic mirror. \newline
	By turning the $\lambda/4$ wave plate in front of the mirror, the trapping light becomes elliptically polarized, defining two orthogonal radial axes with different forces on the particle. The formerly degenerated radial trap frequencies split up in two different peaks in the spectrum. This is necessary for example to cool the particle's motion in the future. For parametric cooling the trapping power is slightly modulated at twice the trapping frequency to reduce the particle's motion, which only works for a known phase. This phase is not uniquely defined for degenerated radial trap frequencies \cite{gieseler_subkelvin_2012}. \newline

\begin{figure}[htbp]
	 	\includegraphics[width=0.48\textwidth]{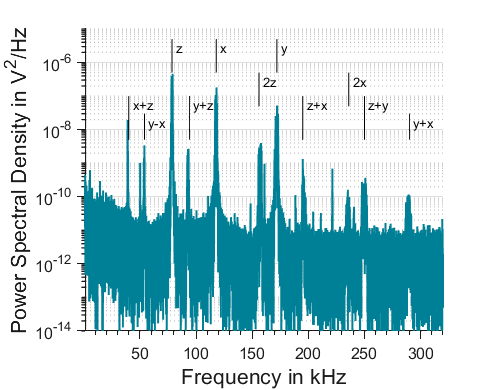}
	 	\caption{\textbf{Power spectral density} of the recorded interference signal for a particle oscillating in an anharmonic region of the optical trap. Fundamental frequencies belong to the three Cartesian axis and are labeled x,y,z accordingly. All other major contributions can be identified as higher harmonics or mixed terms of the fundamentals.}
	 	\label{fig:spectrum}
	 \end{figure}

	Figure \ref{fig:spectrum} shows the resulting spectrum. The most dominant frequencies are fundamental trap oscillations in the three orthogonal directions defined by beam orientation and polarization, where z is aligned with the beam direction, x matches the semi-minor axis and y the semi-major axis of the elliptical polarization. Other peaks can be identified as higher harmonics or mixtures of the main trapping frequencies.

	\subsubsection*{Release- Recapture Experiments}
	 
\begin{figure}[hbt]
            \centering
		\includegraphics[width=0.48\textwidth]{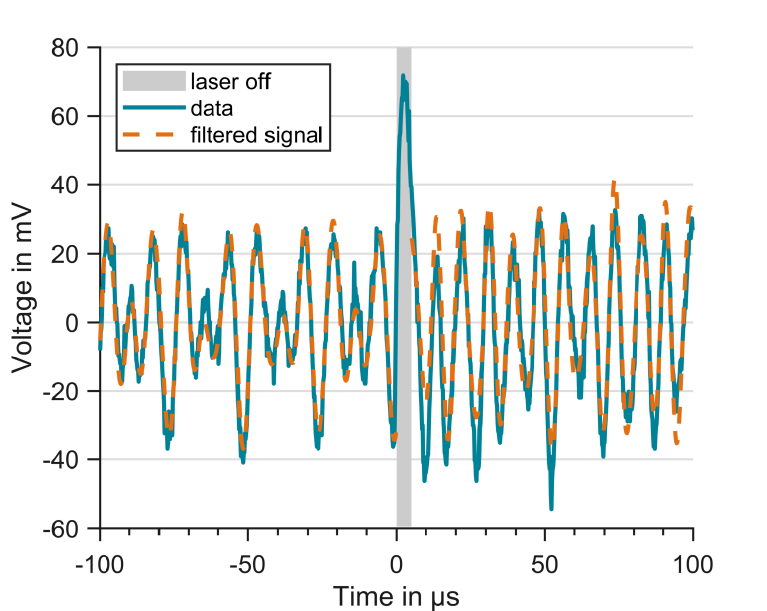}
		\caption{\textbf{Photo diode signal.} Recorded interference signal for a free flight measurement (solid, blue), where the trapping laser is switched off for \SI{7}{\micro\second} at $t=0$ (shaded, gray). The signal is filtered (see chapter methods) to extract the movement in all three directions. Adding the filtered data up gives the dashed orange curve.}
		\label{fig:data}
    \end{figure}

\begin{figure}[h]
			\includegraphics[width=0.48\textwidth]{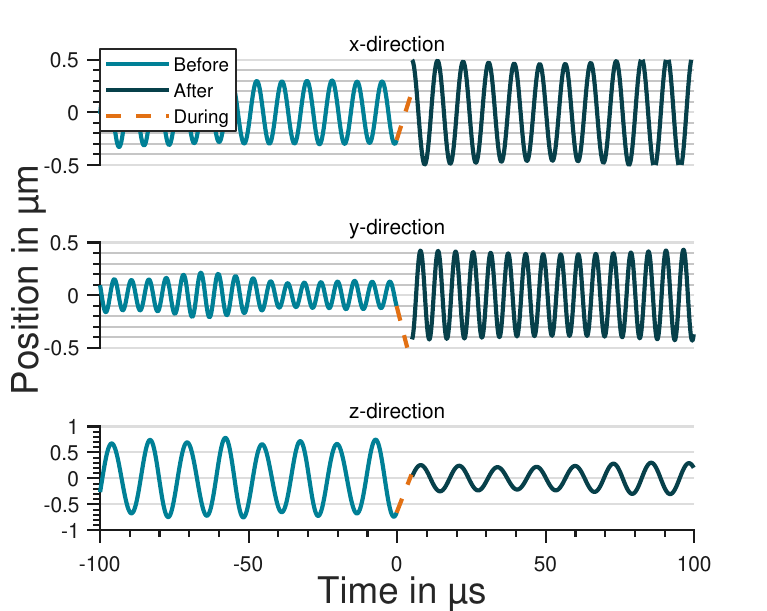}
			\caption{\textbf{Free flight trajectory.} Position data of the particle separated for all three Cartesian axes. Data recorded before (blue) and after (black) the TOF are treated as different data sets. They are connected by the predicted free flight dynamics (orange dashed line) based on position and velocity at the switch off time. }
			\label{fig:traj}
    \end{figure}

	In our free flight experiments in microgravity the particle was released from the optical trap and recaptured. The free flying time is limited to \SI{10}{\micro\second} by the particle's center of mass velocity, which was not artificially reduced for the experiments presented in this paper. An example of such an experimental run is shown in figures \ref{fig:data} and \ref{fig:traj}. \newline  
	Figure \ref{fig:data} shows the recorded voltage over time as solid blue line. At $t=0$ the laser is switched off. Since our trapping laser is used for position detection as well, we have no information about the particle during this time. The rising signal no longer represents the particle position but is an artifact due to the switching process. After \SI{7}{\micro\second} the trapping laser is switched on and one can observe the particle's oscillation again. Due to the free flying time the particle has moved away from the center of the trap, resulting in a larger amplitude of oscillation. The resulting amplitude depends on the particle's velocity, which in turn depends on the oscillation phase, at the time the trap is switched off.\newline
    In order to separate the different movements from each other, as seen in figure \ref{fig:traj}, we apply band pass filters (Butterworth of 10th order), centered around the main frequencies of the signal \cite{hempston_force_2017}. To test for consistency, all three filtered signals are added and compared to the recorded signal with satisfying agreement. The analysis shows the fundamental oscillations (x,y,z) as always present and dominant in our measurements.\newline
    A further movement into anharmonic trapping regions gives rise to mixed frequency components and higher harmonics. While higher harmonics can be fitted, added to the filtered signal and attributed to a single Cartesian coordinate, there is no straightforward procedure for mixed terms that we are aware of. Therefore, to cover movement in the strongly anharmonic regimes of the optical trap it is more adequate to replace the single with a quadrant photo diode to record directional information directly. Our current system is only capable of reliable measurements in the small amplitude (harmonic) regime.\newline
	The interference signal was converted to position information taking advantage of the known kinetic energy of the particle at room temperature \cite{frimmer_controlling_2017}. The calculated position and velocity data can be used to predict the particle's motion while the trap was switched off. We do not expect any significant forces on the particle and therefore assume linear movement from the position and with the velocity in the exact moment the laser is switched off. Switching dynamics in the laser are ignored.\newline
    Figure \ref{fig:traj} displays the particle's position for all three Cartesian axes over time from the same data set seen in Figure \ref{fig:data}. While data before (blue) and after (black) the free flight are treated as separate measurements, the dashed orange line is the predicted free flight. We generally find good agreement between the predicted and measured particle position when the laser is switched back on again. In the same manner this analysis fails for particles with strong anharmonic motion. 	Notice, that the release of a particle does not necessarily lead to an increase in energy, but with the right timing can also lead to a center of mass cooling effect as observed here in z-direction.  

	\section*{Conclusion and Outlook}
	
	We demonstrated the first levitated optomechanics experiment in microgravity. This approach is expected to become important to provide undisturbed free evolution of nanoparticles as quantum mechanical wave packets. Such an observation would be useful for fundamental physics questions as well as a new kind of force sensors based on matter-wave interferometry.  
	Free flights in microgravity of up to \SI{7}{\micro\second} where demonstrated. This time is limited by the particle's velocity. We demonstrated a single photo detector to be sufficient to describe the particles motion in all three translational degrees of freedom, if the movement is restricted to the harmonic and slightly anharmonic region of the trap. These movements can well be described and are consistent with expected free flight trajectories. \newline
    In the next step the setup will be upgraded with an algorithm to provide parametric feedback cooling, which was demonstrated to reach temperatures of a few \si{\milli\kelvin} translating to residual velocities of only hundreds of \si{\micro\meter\per\second} \cite{Vovrosh2017a}. Such improvements would allow to extend the free flying time to the \si{\milli\second} range in our setup. 
	
	\section*{Acknowledgment}
	This project was supported by the German Space Agency (DLR) with funds provided by the Federal Ministry for Economic Affairs and Climate Action (BMWK) due to an enactment of the German Bundestag under Grants No. 50WM2180 (NaiS).\newline
    We would also like to thank Henrik Ulbricht and his working group for their advice on setting up the experiment and for lending us a parabolic mirror in the initial phase of the project. 
	
	\section*{Methods}
	The interferometric signal between scattered light from the particle and reflected light from the mirror is recorded with a single photo diode connected to an oscilloscope (Picoscope, PS 4424A) with a sampling rate of \SI{5}{\mega\sample\per\second}. The information is further processed in MATLAB. A Fourier transformation on data of up to \SI {1}{\second} is used to extract the main frequency components in the signal and determine the fundamental frequencies in all three directions. \newline
	The original datasets with a duration of \SI{10}{\second} are cropped to a total of \SI{200}{\micro\second} around the trapping laser switch off time. The exact switch off is calibrated and post-corrected in time by the strong signal rise observed for every experimental run. For data before the trapping laser switch off, three different 10th order Butterworth filters with half-power frequency values of \SI{5}{\kilo\hertz} to both sides of the resonance are sequentially applied. The results are corrected for filter introduced phase shifts by comparing the output with the original data based on correlation analysis. Since these filters always start a 0 amplitude and need some time to represent the recorded data appropriately, the data set from retrapping the particle was inverted in time before the same set of filters was applied.  \newline

	\section*{Data availability}
	The data supporting the conclusions of this article are available from the
	corresponding author (CV) upon reasonable request.
	
	\section*{Author contributions}
	G.P. was responsible for the daily experimental operation and conducted experiments on ground. G.P. and C.V. conducted experiments in microgravity. C.V. analyzed data and wrote the manuscript which G.P., S.H. and R.B.B. discussed and edited. C.V. and S.H. proposed the experiment, secured funding and coordinated the project. All authors have read and approved the final manuscript.

	\section*{Competing interests}
	All other authors declare no competing interests.

		\bibliographystyle{apsrev4-2}
		\bibliography{FreeFlight}
		
	\end{document}